\documentclass[longauth, letter]{aa} 

\usepackage{graphicx}
\usepackage{comment}
\usepackage[breaklinks,pdfnewwindow=true,colorlinks=true,citecolor=blue]{hyperref}
\usepackage{txfonts}
\usepackage{orcidlink}
\newcommand{\orcid}[1]{\unskip\protect\href{https://orcid.org/#1}{\protect\includegraphics[width=8pt,clip]{logo_orcid}}}

\newcommand{\kms}{km\,s$^{-1}$}

\newcommand{\cms}{cm$^{-2}$}

\newcommand{\msol}{M$_{\odot}$}
\newcommand{\msolyr}{M$_{\odot}$\,yr$^{-1}$}

\begin{document}

   \title{FAUST XVII: Super deuteration in the planet forming system IRS~63 \\
   where the streamer strikes the disk}


    \author{L.~Podio\inst{1}\orcidlink{0000-0003-2733-5372}\and
   {C.~Ceccarelli}\inst{2}\orcidlink{0000-0001-9664-6292}\and
   {C.~Codella}\inst{1,2}\orcidlink{0000-0003-1514-3074}\and
   {G.~Sabatini}\inst{1}\orcidlink{0000-0002-6428-9806}\and
   {D.~Segura-Cox}\inst{3}\orcidlink{0000-0003-3172-6763}\and
   {N.~Balucani}\inst{4}\orcidlink{0000-0001-5121-5683}\and
   {A.~Rimola}\inst{5}\orcidlink{0000-0002-9637-4554}\and
   {P.~Ugliengo}\inst{6}\orcidlink{0000-0001-8886-9832}\and
   {C.~J.~Chandler}\inst{7}\orcidlink{0000-0002-7570-5596}\and
   {N.~Sakai}\inst{8}\orcidlink{0000-0002-3297-4497}\and
   {B.~Svoboda}\inst{7}\orcidlink{0000-0002-8502-6431}\and
   {J.~Pineda}\inst{9}\orcidlink{0000-0002-3972-1978}\and
   {M.~De Simone}\inst{1,10}\orcidlink{0000-0001-5659-0140}\and
   {E.~Bianchi}\inst{11}\orcidlink{0000-0001-9249-7082}\and
   {P.~Caselli}\inst{9}\orcidlink{0000-0003-1481-7911}\and
   {A.~Isella}\inst{12}\orcidlink{0000-0001-8061-2207}\and
   {Y.~Aikawa}\inst{13}\orcidlink{0000-0003-3283-6884}\and
   {M.~Bouvier}\inst{14}\orcidlink{0000-0003-0167-0746}\and
   {E.~Caux}\inst{15}\orcidlink{0000-0002-4463-6663}\and
   {L.~Chahine}\inst{2}\orcidlink{0000-0003-3364-5094}\and
   {S.~B.~Charnley}\inst{16}\orcidlink{0000-0001-6752-5109}\and
   {N.~Cuello}\inst{2}\orcidlink{0000-0003-3713-8073}\and
   {F.~Dulieu}\inst{17}\orcidlink{0000-0001-6981-0421}\and
   {L.~Evans}\inst{18}\orcidlink{0009-0006-1929-3896}\and
   {D.~Fedele}\inst{1}\orcidlink{0000-0001-6156-0034}\and
   {S.~Feng}\inst{19}\orcidlink{0000-0002-4707-8409}\and
   {F.~Fontani}\inst{1,9,31}\orcidlink{0000-0003-0348-3418}\and
   {T.~Hama}\inst{20,21}\orcidlink{0000-0002-4991-4044}\and
   {T.~Hanawa}\inst{22}\orcidlink{0000-0002-7538-581X}\and
   {E.~Herbst}\inst{23}\orcidlink{0000-0002-4649-2536}\and
   {T.~Hirota}\inst{24}\orcidlink{0000-0003-1659-095X}\and
   {I.~Jim\'{e}nez-Serra}\inst{25}\orcidlink{0000-0003-4493-8714}\and
   {D.~Johnstone}\inst{26,27}\orcidlink{0000-0002-6773-459X}\and
   {B.~Lefloch}\inst{28}\orcidlink{0000-0002-9397-3826}\and
   {R.~Le Gal}\inst{2,29}\orcidlink{0000-0003-1837-3772}\and
   {L.~Loinard}\inst{30}\orcidlink{0000-0002-5635-3345}\and
   {H.~Baobab Liu}\inst{32}\orcidlink{0000-0003-2300-2626}\and
   {A.~L\'{o}pez-Sepulcre}\inst{2,29}\orcidlink{0000-0002-6729-3640}\and
   {L.~T.~Maud}\inst{10}\orcidlink{0000-0002-7675-3565}\and
   {M.~J.~Maureira}\inst{9}\orcidlink{0000-0002-7026-8163}\and
   {F.~Menard}\inst{2}\orcidlink{0000-0002-1637-7393}\and
   {A.~Miotello}\inst{10}\orcidlink{0000-0002-7997-2528}\and
   {G.~Moellenbrock}\inst{7}\orcidlink{0000-0002-3296-8134}\and
   {H.~Nomura}\inst{33}\orcidlink{0000-0002-7058-7682}\and
   {Y.~Oba}\inst{34}\orcidlink{0000-0002-6852-3604}\and
   {S.~Ohashi}\inst{8}\orcidlink{0000-0002-9661-7958}\and
   {Y.~Okoda}\inst{8}\orcidlink{0000-0003-3655-5270}\and
   {Y.~Oya}\inst{35,36}\orcidlink{0000-0002-0197-8751}\and
   {T.~Sakai}\inst{37}\orcidlink{0000-0003-4521-7492}\and
   {Y.~Shirley}\inst{38}\orcidlink{0000-0002-0133-8973}\and
   {L.~Testi}\inst{1,39}\orcidlink{0000-0003-1859-3070}\and
   {C.~Vastel}\inst{15}\orcidlink{0000-0001-8211-6469}\and
   {S.~Viti}\inst{14}\orcidlink{0000-0001-8504-8844}\and
   {N.~Watanabe}\inst{34}\orcidlink{0000-0001-8408-2872}\and
   {Y.~Watanabe}\inst{40}\orcidlink{0000-0002-9668-3592}\and
   {Y.~Zhang}\inst{8}\orcidlink{0000-0001-7511-0034}\and
   {Z.~E.~Zhang}\inst{41}\orcidlink{0000-0002-9927-2705}\and
   {S.~Yamamoto}\inst{42}\orcidlink{0000-0002-9865-0970}   
   }
   
      \institute{INAF, Osservatorio Astrofisico di Arcetri, Largo E. Fermi 5, I-50125, Firenze, Italy; \email{linda.podio@inaf.it} 
      \and Univ. Grenoble Alpes, CNRS, IPAG, 38000 Grenoble, France
      \and Department of Astronomy, The University of Texas at Austin, 2515 Speedway, Austin, Texas 78712, USA
      \and Department of Chemistry, Biology, and Biotechnology, The University of Perugia, Via Elce di Sotto 8, 06123 Perugia, Italy
      \and Departament de Qu\'{i}mica, Universitat Aut\`{o}noma de Barcelona, 08193 Bellaterra, Spain
      \and Dipartimento di Chimica and Nanostructured Interfaces and Surfaces (NIS) Centre, Universit\`{a} degli Studi di Torino, via P. Giuria 7, I-10125 Torino, Italy
      \and National Radio Astronomy Observatory, PO Box O, Socorro, NM 87801, USA
      \and RIKEN Cluster for Pioneering Research, 2-1, Hirosawa, Wako-shi, Saitama 351-0198, Japan
      \and Center for Astrochemical Studies, Max-Planck-Institut f\"{u}r extraterrestrische Physik (MPE), Gie$\beta$enbachstr. 1, D-85741 Garching, Germany
      \and European Southern Observatory, Karl-Schwarzschild Str. 2, 85748 Garching bei M\"{u}nchen, Germany
      \and Excellence Cluster ORIGINS, Boltzmannstraße 2, D-85748 Garching bei M\"{u}nchen, Germany
      \and Department of Physics and Astronomy, Rice University, 6100 Main Street, MS-108, Houston, TX 77005, USA
      \and Department of Astronomy, The University of Tokyo, 7-3-1 Hongo, Bunkyo-ku, Tokyo 113-0033, Japan
      \and Leiden Observatory, Leiden University, P.O. Box 9513, 2300 RA Leiden, The Netherlands
      \and IRAP, Univ. de Toulouse, CNRS, CNES, UPS, Toulouse, France
      \and Astrochemistry Laboratory, Code 691, NASA Goddard Space Flight Center, 8800 Greenbelt Road, Greenbelt, MD 20771, USA
      \and CY Cergy Paris Universit\'{e}, Sorbonne Universit\'{e}, Observatoire de Paris, PSL University, CNRS, LERMA, F-95000, Cergy, France
      \and School of Physics and Astronomy, University of Leeds, Leeds LS2 9JT, UK
      \and Department of Astronomy, Xiamen University, Xiamen, Fujian 361005, P. R. China
      \and Komaba Institute for Science, The University of Tokyo, 3-8-1 Komaba, Meguro, Tokyo 153-8902, Japan
      \and Department of Basic Science, The University of Tokyo, 3-8-1 Komaba, Meguro, Tokyo 153-8902, Japan
      \and Center for Frontier Science, Chiba University, 1-33 Yayoi-cho, Inage-ku, Chiba 263-8522, Japan
      \and Department of Chemistry, University of Virginia, McCormick Road, PO Box 400319, Charlottesville, VA 22904, USA
      \and National Astronomical Observatory of Japan, Osawa 2-21-1, Mitaka-shi, Tokyo 181-8588, Japan
      \and Centro de Astrobiolog\'{i}a (CAB), CSIC-INTA, Ctra. de Torrej\'{o}n a Ajalvir, km 4, 28850, Torrej\'{o}n de Ardoz, Spain
      \and NRC Herzberg Astronomy and Astrophysics, 5071 West Saanich Road, Victoria, BC, V9E 2E7, Canada
      \and Department of Physics and Astronomy, University of Victoria, Victoria, BC, V8P 5C2, Canada
      \and Universit\'{e} de Bordeaux – CNRS Laboratoire d’Astrophysique de Bordeaux, 33600 Pessac, France
      \and Institut de Radioastronomie Millim\'{e}trique, 38406 Saint-Martin d’H\`{e}res, France
      \and Instituto de Radioastronom\'{i}a y Astrof\'{i}sica , Universidad Nacional Aut\'{o}noma de M\'{e}xico, A.P. 3-72 (Xangari), 8701, Morelia, Mexico
      \and LERMA, Observatoire de Paris, PSL Research University, CNRS, Sorbonne Universit{\'e}, 92190 Meudon, France
      \and Institute of Astronomy and Astrophysics, Academia Sinica, 11F of Astronomy-Mathematics Building, AS/NTU No.1, Sec. 4, Roosevelt Rd., Taipei 10617, Taiwan, R.O.C.
      \and Division of Science, National Astronomical Observatory of Japan, 2-21-1 Osawa, Mitaka, Tokyo 181-8588, Japan
      \and Institute of Low Temperature Science, Hokkaido University, N19W8, Kita-ku, Sapporo, Hokkaido 060-0819, Japan
      \and Department of Physics, The University of Tokyo, 7-3-1, Hongo, Bunkyo-ku, Tokyo 113-0033, Japan
      \and Yukawa Institute for Theoretical Physics, Kyoto Univ. Oiwake-cho, Kitashirakawa, Sakyo-ku, Kyoto-shi, Kyoto-fu 606-8502, Japan
      \and Graduate School of Informatics and Engineering, The University of Electro-Communications, Chofu, Tokyo 182-8585, Japan
      \and Steward Observatory, 933 N Cherry Ave., Tucson, AZ 85721 USA
      \and Dipartimento di Fisica e Astronomia “Augusto Righi” Viale Berti Pichat 6/2, Bologna, Italy
      \and Materials Science and Engineering, College of Engineering, Shibaura Institute of Technology, 3-7-5 Toyosu, Koto-ku, Tokyo 135-8548, Japan
      \and Star and Planet Formation Laboratory, RIKEN Cluster for Pioneering Research, Wako, Saitama 351-0198, Japan
      \and SOKENDAI (The Graduate University for Advanced Studies), Shonan Village, Hayama, Kanagawa 240-0193, Japan
      \and 
      }

   \date{Received ; accepted }

 
  \abstract
   {Recent observations suggest that planets formation starts early, in protostellar disks of $\le10^5$ yrs, which are characterized by strong interactions with the environment, e.g., through accretion streamers and molecular outflows. }
   {To investigate the impact of such phenomena on disk physical and chemical properties it is key to understand what chemistry planets inherit from their natal environment.} 
   {In the context of the ALMA Large Program (LP) Fifty AU STudy of the chemistry in the disk/envelope system of Solar-like protostars (FAUST), we present observations on scales from $\sim$1500~au to $\sim$60~au of H$_2$CO, HDCO, and D$_2$CO towards the young planet-forming disk IRS~63.}
   {H$_2$CO probes the gas in the disk as well as in a large scale streamer ($\sim 1500$ au) impacting onto the South-East (SE) disk side. We detect for the first time deuterated formaldehyde, HDCO and D$_2$CO, in a planet-forming disk, and HDCO in the streamer that is feeding it. This allows us to estimate the deuterium fractionation of H$_2$CO in the disk: [HDCO]/[H$_2$CO]$\sim 0.1-0.3$ and [D$_2$CO]/[H$_2$CO]$\sim 0.1$. 
   Interestingly, while HDCO follows the H$_2$CO distribution in the disk and in the streamer, the distribution of D$_2$CO is highly asymmetric, with a peak of the emission (and [D]/[H]  ratio) in the SE disk side, where the streamer crashes onto the disk. In addition, D$_2$CO is detected in two spots along the blue- and red-shifted outflow. 
   This suggests that: (i) in the disk, HDCO formation is dominated by gas-phase reactions similarly to H$_2$CO, while (ii) D$_2$CO was mainly formed on the grain mantles during the prestellar phase and/or in the disk itself, and is at present released in the gas-phase in the shocks driven by the streamer and the outflow.  
   }
   {These findings testify on the key role of streamers in the build-up of the disk both concerning the final mass available for planet formation and its chemical composition.}


   \maketitle
%

\section{Introduction}

\begin{table*}
	\setlength{\tabcolsep}{50pt}
	\renewcommand{\arraystretch}{0.9}
	\renewcommand{\tabcolsep}{5pt}
\centering
  \caption[]{\label{tab:lines} Properties of the observed lines from the Cologne Database for Molecular Spectroscopy (CDMS, \citealt{Mueller2005}) and of the obtained line cubes. The last two columns report the line integrated intensities on the SE and NW disk sides.}
  \begin{tabular}[h]{ccccccccc}
    \hline
    \hline
 Species & Transition & Frequency & $E_{\rm up}$ & Clean beam & Channel & r.m.s. & $I_{\rm SE-disk}$ &  $I_{\rm NW-disk}$ \\ 
         &            & (MHz)     & (K)          & ($\arcsec\times\arcsec$) ($\degr$) & (km s$^{-1}$) & (mJy beam$^{-1}$) & \multicolumn{2}{c}{(mJy \kms)} \\
    \hline 
p-H$_2$CO & $3_{0,3}-2_{0,2}$ & 218222.192 & 21 & $0.52\times0.42$ ($-87.3\degr$) & 0.17 & 2.5 & 24.6 & 26.4 \\ 
%
o-D$_2$CO & $4_{0,4}-3_{0,3}$ & 231410.234 & 28 & $0.49\times0.40$ ($-86.6\degr$) & 0.16 & 2.5 & 10.9 & $5$ \\ 
o-D$_2$CO & $4_{2,3}-3_{2,2}$ & 233650.441 & 50 & $0.48\times0.39$ ($-86.1\degr$) & 0.63 & 1.0 &  $<3$ & $<3$ \\ 
%
HDCO    & $4_{1,4}-3_{1,3}$ & 246924.600 & 38 & $0.53\times0.38$ ($-84.9\degr$) & 1.19 & 1.0 & 10.6 & 9.4 \\ 
HDCO    & $4_{2,2}-3_{2,1}$ & 259034.910 & 63 & $0.50\times0.37$ ($-84.5\degr$) & 0.14 & 3.5 & 5.8  & 4.5 \\ 
    \hline     
  \end{tabular}
\end{table*}

%
Observations of embedded Class 0 and I sources provide a number of evidence that planets formation starts early, in disks with ages of $\le 10^5$ years: evidence of dust growth up to mm sizes in young disks \citep[e.g., ][]{miotello14a,harsono18,sabatini24}, as well as observations of rings and gaps in their dust distribution strongly support this hypothesis \citep[e.g., ][]{alma15,sheehan18,segura-cox20}. 

These young disks are characterized by strong interactions with the surrounding environment, which may affect the disk's physical (e.g. mass and kinematics), and chemical (e.g., molecular abundances) properties. For example, recent observations at millimeter wavelengths revealed streamers channeling gas and dust from the outer regions of the envelope or even from the surrounding cloud to the disk, hence 
playing a key role in regulating the mass of the disk 
\citep[e.g., ][]{pineda20,pineda23,valdivia-mena22,cacciapuoti24,mercimek23}
as well as its chemical composition \citep[e.g., ][]{garufi22,maureira22,valdivia-mena23,bianchi22a,bianchi23b,codella24}.
It is therefore crucial to investigate how the properties of young disks are affected by environmental processes to understand how planets formed in the Solar System, as well as in other planetary systems, and what chemistry they inherited from their natal environment. 

To this aim, the Large Program FAUST is designed to observe young disks  with the Atacama Large Millimeter/submillimeter Array (ALMA) using multiple antennas configurations which allows us to probe the chemistry from large (thousands of au) to small (50 au) scales \citep{codella21}.
In the sample of protostars observed by FAUST, an ideal candidate to probe the disk properties and their interconnection with the environment is the Class I source IRS~63, located in the nearby Ophiuchus star forming region at a distance of 144 pc \citep{ortiz-leon18a}.
ALMA observations of mm continuum emission at $\sim 5$ au resolution have shown a relatively large dusty disk ($\sim 82$ au) characterized by multiple annular substructures (two rings and two gaps), suggestive of ongoing planet formation \citep{segura-cox20}. Recent observations of $^{13}$CO and C$^{18}$O 
by \citet{flores23}
revealed the gaseous disk which extends out to $\sim 260$ au (i.e. 3-4 times larger than the dust disk),  and enabled an estimation of 
the protostellar mass from keplerian velocity fitting $M_{*} = 0.5 \pm 0.2$ \msol. 
Moreover, their observations show that IRS~63 is associated to a  bipolar outflow (seen in $^{12}$CO), a rotating envelope (in $^{13}$CO), and a large scale streamer connecting the envelope to the disk (in C$^{18}$O) with small-scale spiral structures at the edge of the dust continuum (in SO).
They show that the streamer and the spiral structures  originate from an infalling rotating structure that feeds the young protostellar disk at a  rate of $\sim 10^{-6}$ \msolyr, i.e. two orders of magnitude larger than the disk-to-star mass accretion rate of $\sim 10^{-8}$ \msolyr\, \citep{flores23}. 

%

In this Letter, we analyse for the first time the deuterated isotopologues of H$_2$CO towards the planet-forming disk IRS~63 and its streamer. 
Theoretical and observational studies indicate that deuterated molecules are  "fossils" of the prestellar core phase, and their abundances may be inherited from these early phases \citep[e.g., ][]{caselli12a,ceccarelli14,jensen21}. 
Hence, they are ideal to investigate  what is the impact, if any, of the large scale streamer 
on the  disk chemical content at the time of planet formation.

\begin{figure*}[ht]
    \centering
    \includegraphics[width=\linewidth]{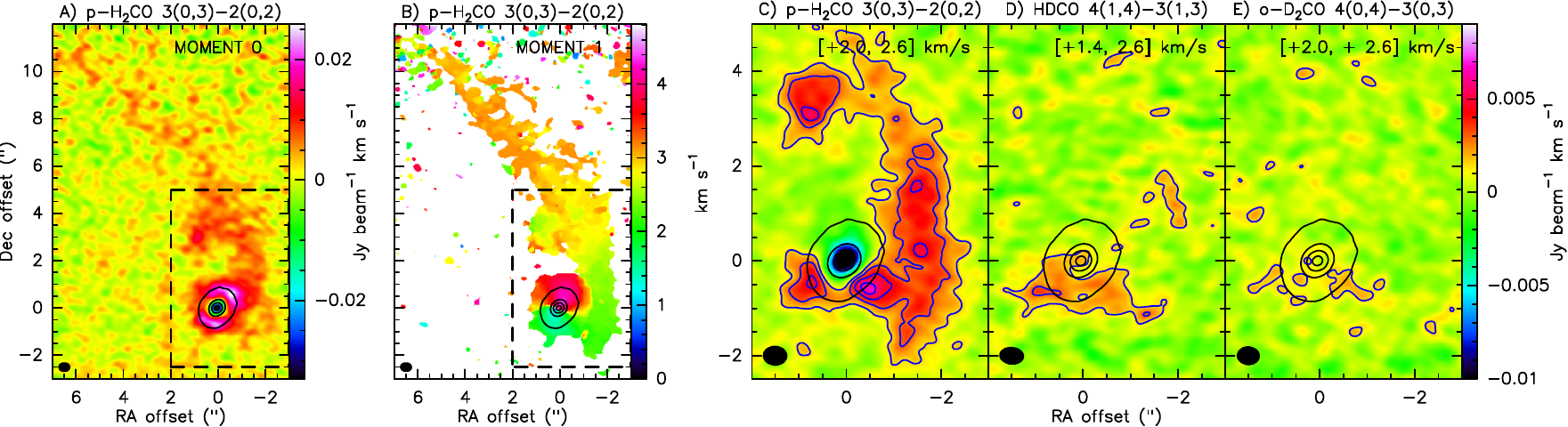}
    \caption{The streamer and the disk of IRS~63 as probed by p-H$_2$CO $3(0,3)-2(0,2)$, HDCO $4(1,4)-3(1,3)$, and o-D$_2$CO $4(0,4)-3(0,3)$. {\it From left to right:} Panels A) and B) show the moment 0 and moment 1 maps of the H$_2$CO emission. Panels C), D), and E) show a zoom-in of H$_2$CO, HDCO and D$_2$CO integrated on the blue-shifted velocity range [+2.0, +2.6] \kms, which probes the terminal portion of the streamer impacting onto the disk. The map of HDCO covers a larger velocity range [+1.4, +2.6] \kms, due to the larger channel width of the line cube (1.19 \kms). First contours and steps correspond to $3\sigma$ (0.5 mJy beam$^{-1}$ \kms).}
    \label{fig:streamer-all}
\end{figure*}

\begin{figure*}[ht]
    \centering
    \includegraphics[width=0.8\linewidth]{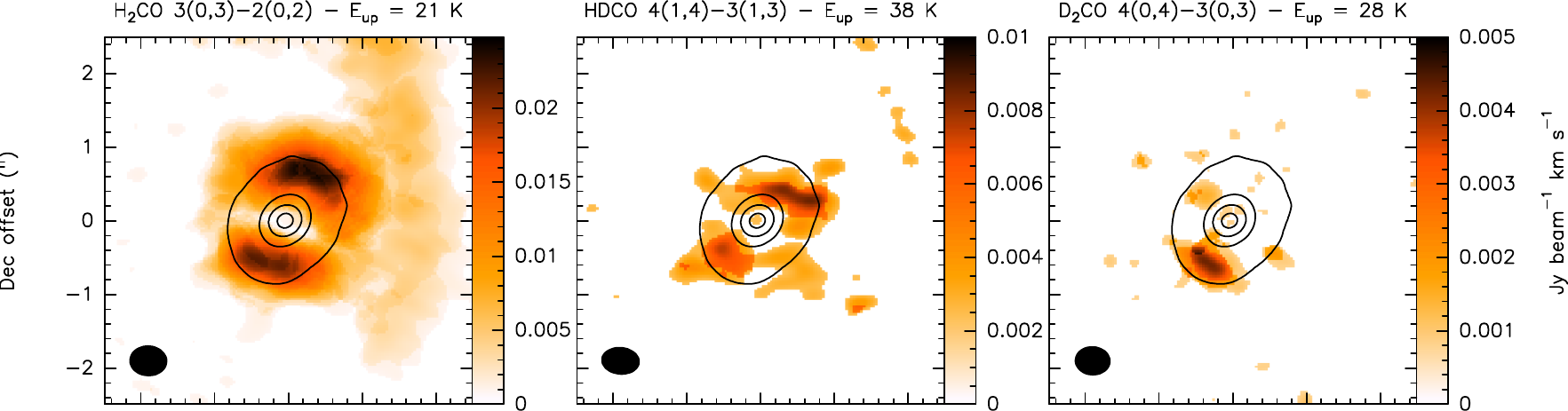}
    \includegraphics[width=0.8\linewidth]{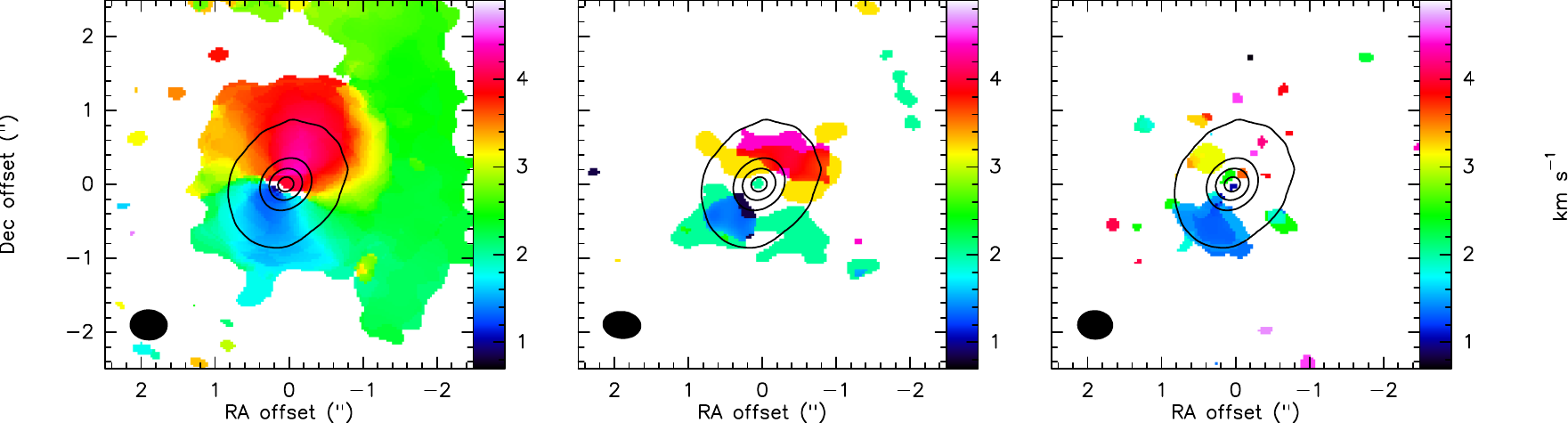}
    \caption{Zoom-in of the integrated intensity (moment 0, {\it upper panels}) and intensity weighted velocity (moment 1, {\it lower panels}) maps of the H$_2$CO, HDCO, and D$_2$CO line emission towards the IRS 63 disk (same transitions as in Fig. \ref{fig:streamer-all}). The  continuum emission at 1.3~mm is in black (contours from $10\sigma$ with steps of $500\sigma$, with $\sigma = 0.7$ mJy\,beam$^{-1}$). The color scale is in Jy \kms\,beam$^{-1}$ and in \kms\, for the moment 0 and moment 1 maps, respectively. In each panel the transition and upper level energy are labeled on the top, the beam on the lower-left.}
    \label{fig:mom}
\end{figure*}

\section{Observations and data reduction}
In the context of the ALMA LP FAUST (2018.1.01205.L, PI: S. Yamamoto), the Class I protostellar disk IRS~63 was observed between 2018 and 2020, in Band 6, using three configurations: two configurations with the main array (12 m, C43-4 and C43-1), and one configuration with the Atacama Compact Array (ACA, 7 m antennas). The observations were taken using two spectral setups, each covering 12 narrow spectral windows (SPW) centered on selected molecular lines (channel width 0.12 MHz, coverage $\sim 58$ MHz), and one wide SPW which covers $\sim 2$~GHz with a channel width of $0.49$ MHz.
The spectral setups cover p-H$_2$CO $3_{0,3}-2_{0,2}$ ($E_{\rm up} \sim 21$ K), o-D$_2$CO $4_{0,4}-3_{0,3}$, $4_{2,3}-3_{2,2}$ ($E_{\rm up} \sim 28, 50$ K), and HDCO $4_{1,4}-3_{1,3}$, $4_{2,2}-3_{2,1}$ ($E_{\rm up} \sim 38, 63$ K). The $4_{2,3}-3_{2,2}$ and HDCO $4_{1,4}-3_{1,3}$ lines are covered by the wide SPW, hence are at lower spectral resolution. The observations taken in the three configurations were combined using the ‘mosaic’ gridder in {\it tclean} within the Common Astronomy Software Applications package (CASA). The line-free continuum channels have been identified and subtracted and the data have been self-calibrated, imaged and cleaned applying robust weighting (robust 0.5), obtaining a clean beam of $\sim 0\farcs45$ for the lines.
The properties of the observed lines (transition, frequency, and upper level energy, $E_{\rm up}$), and of the obtained line cubes (clean beam, channel width, and root mean square (r.m.s.) noise) are summarized in Table \ref{tab:lines}.
The channel maps of the H$_2$CO line cube, and of the brightest HDCO and D$_2$CO line cubes are shown in Figs.~\ref{fig:h2co_streamer_channels}, \ref{fig:h2co_channels}, \ref{fig:hdco_channels}, and \ref{fig:d2co_channels}.
Finally, integrated intensity maps (moment 0) and intensity-weighted velocity maps (moment 1) are obtained by 
applying a $2\sigma$ clipping (Fig. \ref{fig:mom}). 

\section{Results}

The moment 0 and moment 1 maps of formaldehyde (H$_2$CO), and of the  lower $E_{\rm up}$ lines  of its singly and doubly deuterated  forms (i.e. HDCO 246.9 GHz and D$_2$CO 231.4 GHz), are shown in Figs.~\ref{fig:streamer-all} and \ref{fig:mom}, with overlapped the contours of the continuum emission at 1.3~mm (see Sect. \ref{sect:continuum}).
Fig. \ref{fig:streamer-all} shows a $10"\times15"$ region to investigate the emission on large scales (see Sect. \ref{sect:streamer}), while Fig. \ref{fig:mom} shows a zoom-in ($4"\times4"$) focusing on the molecules distribution in the disk (see Sect. \ref{sect:disk}).


\subsection{Continuum emission at 1.3~mm: the dusty disk}
\label{sect:continuum}

The continuum emission shows a smooth distribution, with no signatures of the rings and gaps first identified by \citet{segura-cox20}, due to the $\sim 10$ times lower angular resolution of the FAUST map ($0\farcs38 \times 0\farcs30$, PA$=84.25\degr$) compared to their map ($0\farcs05 \times 0\farcs03$, i.e. $\sim 7 \times4$ au).
A 2D Gaussian fit is performed to estimate the disk parameters: peak coordinates (R.A.(J2000) $= 16^h 31^m 35^s.66$, Dec(J2000) $=-24\degr 01' 29\farcs97$), peak intensity ($F_{\rm peak} = 112.7 \pm 0.4$ mJy/beam), and integrated flux density at 1.3~mm ($293 \pm 1$ mJy), disk position angle and inclination (PA$_{\rm disk} = 149 \pm 1 \degr$, $i=47\pm1 \degr$). The values are in agreement with those estimated by \citet{segura-cox20} from their higher resolution continuum maps.

\subsection{The streamer feeding the IRS~63 disk}
\label{sect:streamer}

The moment 0 and moment 1 maps  in Fig.~\ref{fig:streamer-all} show that the H$_2$CO line at 218.2 GHz probes the gas in the disk of IRS~63, as well as in the accretion streamer first identified in C$^{18}$O by \citet{flores23}.
The streamer extends from $\sim 1500$ au North-East (NE) to IRS~63 down to the western disk side, and shows a velocity gradient with  velocities spanning from $\sim 3.3$ \kms, i.e. red-shifted with respect to systemic ($V_{\rm LSR} \sim +2.8$ \kms) in the northern part, down to $\sim 2$ \kms\, in the southern part, where the streamer appears to connect to the disk.
The channel maps in Fig. \ref{fig:h2co_streamer_channels} show that the spatial distribution of the H$_2$CO emission in the disk, in each velocity channel, is not symmetric with respect to the disk major axis, as expected in a Keplerian rotating disk. This asymmetry is likely produced by the perturbation induced by the streamer as it approaches the disk. 
A detailed analysis of the physical, kinematical, and chemical properties of the streamer is beyond the scope of this letter and will be presented in a paper by Pineda et al. (in prep).
In the present study we focus on the chemistry of deuterated H$_2$CO in IRS~63.
Panels C), D), and E) of Fig. \ref{fig:streamer-all} show a zoom-in of the H$_2$CO 218.2 GHz, HDCO 246.9 GHz, and D$_2$CO 231.4 GHz line emission integrated on the blue-shifted velocity range where the streamer emits, i.e. $V_{\rm LSR} =[+2.0, +2.6]$ \kms.
The H$_2$CO map shows that the terminal blue-shifted part of the streamer connects to the disk perpendicularly to its major axis and towards the SE disk side. 
Interestingly,  small selected portions of the streamer are also detected in HDCO and D$_2$CO, more specifically:
(i) the brightest part of the arc located NW to the source at an offset of $0\farcs5 - 2"$ in Dec and $-1\farcs5$ and $-2"$ in R.A. is also detected in HDCO; and
(ii)
the terminal portion of the streamer impacting onto the SE disk side, is also detected in the  HDCO and D$_2$CO lines.  

\subsection{{\rm H$_2$CO, HDCO} and {\rm D$_2$CO} in the disk and outflow}
\label{sect:disk}

Fig.~\ref{fig:mom} shows a zoom-in of the moment 0 and 1 maps on the IRS~63 disk. The  H$_2$CO line at 218.2 GHz probes the gas in the disk, which extends beyond the continuum emission from the mm dusty grains, as typically observed in disks \citep[e.g., ][]{garufi21,oberg21}, and shows the typical velocity pattern of the gas in a rotating disk.
A narrow stripe with no emission is observed along the minor axis due to absorption of the line emission by H$_2$CO molecules in the surrounding envelope at velocities close to systemic ($V_{\rm LSR} \sim  +2.8$ km\,s$^{-1}$).
Moreover, the channel maps of H$_2$CO shown in the Appendix (Fig. \ref{fig:h2co_channels}) show negative fluxes in the inner $\sim 20$ au disk region, in the channels close to systemic velocity (between 2.15 and 3.15 \kms), due to optically thick continuum emission and absorption by molecules along the line of sight. The emission detected west to the disk probes the large scale streamer discussed in Sect. \ref{sect:streamer}.

The HDCO $4_{1,4}-3_{1,3}$ line ($E_{\rm up} \sim 38$ K) shows a spatial and velocity distribution similar to H$_2$CO, with two bright spots of emission in the NW and SE disk sides along the disk major axis.
The emission in the H$_2$CO and HDCO lines is brighter in the north-west side of the disk, where also the continuum has higher intensity values at most radii compared to the south-east side of the disk \citep[][see Figs. \ref{fig:mom} and \ref{fig:hdco_channels}]{segura-cox20}.
This indicates that the considered H$_2$CO and HDCO lines probe roughly the same disk region despite the upper level energy of the HDCO line is slightly higher than for H$_2$CO ($E_{\rm up} \sim 38$ K, and $\sim 21$ K, respectively).
Instead, the D$_2$CO $4_{0,4}-3_{0,3}$ line at 231.4 GHz ($E_{\rm up} \sim 28$ K) is firmly detected in the south-east disk side, while no emission above the $2\sigma$ threshold is detected in the NW side (see Figs. \ref{fig:mom} and \ref{fig:d2co_channels}).
The fact that, despite the similar $E_{\rm up}$, the spatial distribution of D$_2$CO is different from that of HDCO and H$_2$CO, suggests a different chemical origin (see Sect.~\ref{sect:discussion}).
The moment 0  map of HDCO 246.9 GHz and  D$_2$CO 231.4 GHz lines also show two spots of emission along the disk minor axis, i.e. along the direction of  the bipolar outflow identified in $^{12}$CO by \citet{flores23}. The moment 1 map shows that the velocity gradient is consistent with that of the $^{12}$CO outflow, with blue-shifted emission towards the South-West and red-shifted towards the North-East.
Finally, as already pointed out in Sect.\ref{sect:streamer}, the moment 0 and 1 map of HDCO also shows emission in a small portion of the arc-like streamer located NW to the disk identified in H$_2$CO . 

The higher excitation HDCO 259.0 GHz and D$_2$CO 233.6 GHz lines ($E_{\rm up} \sim 63$ K and $50$ K, respectively) are fainter and no emission is detected in the moment 0 maps obtained applying a $2\sigma$ threshold. 
However, the fainter HDCO line at 259.0 GHz is detected in the disk-integrated spectra extracted integrating out to a radius of $1\farcs3$, i.e. 187 au, while no emission is seen in the higher excitation D$_2$CO line at 233.7 GHz (see Fig. \ref{fig:spec}). 


\begin{figure*}[ht]
    \centering
    \includegraphics[width=0.8\linewidth]{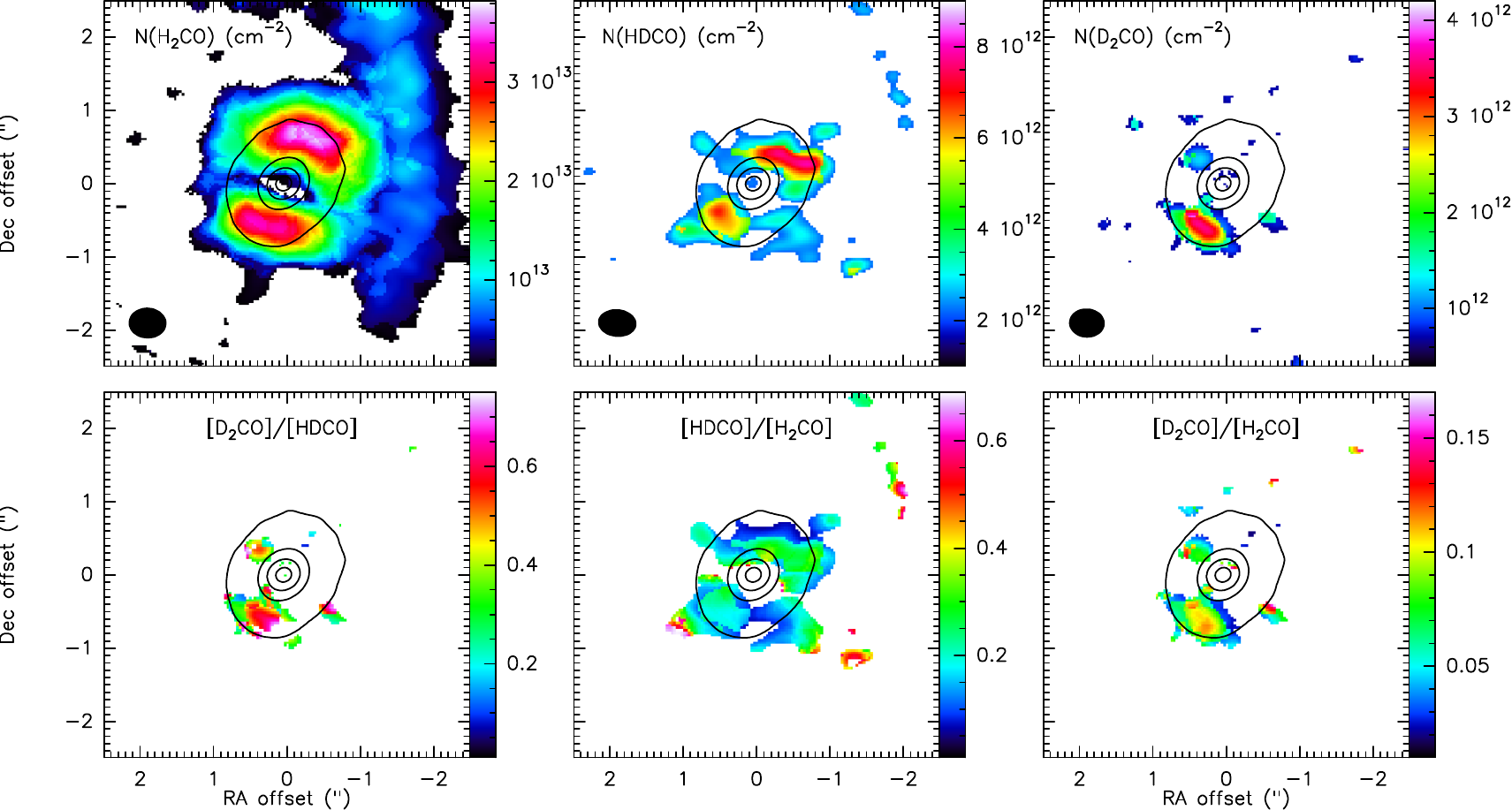}
    \caption{{\it Upper panels:} column density maps of H$_2$CO, HDCO, and D$_2$CO, derived assuming LTE, optically thin emission at the temperature derived from the HDCO line ratio ($T_{ex} =45$ K). {\it Lower panels:}) Maps of the abundance ratios: [D$_2$CO]/[HDCO], [HDCO]/[H$_2$CO], [D$_2$CO]/[H$_2$CO]. The  continuum emission at 1.3~mm is in black (contours from $10\sigma$ with steps of $500\sigma$, with $\sigma = 0.7$ mJy\,beam$^{-1}$).}
    \label{fig:ratios}
\end{figure*}

\subsection{Molecular column density and deuterium fractionation}

Figure  \ref{fig:spec} shows the H$_2$CO, HDCO, and D$_2$CO line spectra obtained by integrating the line cubes on the SE and NW disk sides, on a $0\farcs35$ area (corresponding to $\sim 1.4$ beams) where HDCO emission is detected.  
This allows us to verify that the detected H$_2$CO 218.2 GHz, HDCO 246.9 GHz, and D$_2$CO 231.4 GHz lines originate from the same disk layer, and to recover the emission in the fainter HDCO 259.0 GHz and D$_2$CO 233.7 GHz lines (undetected in the channel and moment 0 maps). 
The integrated spectra show that the fainter, higher excitation, HDCO line at 259.0 GHz is  detected at $>2\sigma$ both in the SE and in the NW disk side, while D$_2$CO is detected only in the lower excitation line at $E_{\rm up}=28$ K.
The peak intensity of the H$_2$CO, HDCO, and D$_2$CO lines in the  SE and NW disk sides is $\sim 1.5-2$ K (H$_2$CO), and $\sim 0.2-0.6$ K (HDCO, and D$_2$CO). If we assume that the lines are in Local Thermodynamic Equilibrium (LTE), i.e. the gas density is larger than the lines critical density, which is very likely in the disk, the line brightness temperature $T_{\rm mb}$, in K, corresponds to the gas temperature if the lines are optically thick. The low  $T_{\rm mb}$ values in the spectra ($<2$ K) indicates that the lines are optically thin.  
The integrated intensities and upper limits in the SE and NW disk sides are reported in Table \ref{tab:lines}.

The molecular column densities and gas temperature are obtained in two steps, under the assumption of LTE and optically thin emission.
First, the ratio of the integrated intensities of the two HDCO lines is used to estimate the excitation temperature.
Even though the range of upper level energies covered by the lines is small ($E_{\rm up}=38$ K and $63$ K),  the HDCO line ratios indicate that the excitation temperature, $T_{\rm ex}$, is $\sim49$ K in the SE disk side, where the streamer impacts onto the disk, and $T_{\rm ex} \sim 39$ K in the NW disk side.
Second, we assume that the H$_2$CO, HDCO, and D$_2$CO emission originates from the same disk layer (as indicated by the fact that they are co-spatial in the moment 0 maps and emit on the same velocity range in the integrated spectra) at an average excitation temperature of $45$ K, and obtain their column density, $N$, pixel by pixel from the line integrated intensities (moment 0) maps\footnote{Partition functions are taken from the CDMS \citep{Mueller2005} and takes into account a statistical ortho-to-para ratio of 3 for H$_2$CO, and 2 for D$_2$CO.}.
Fig. \ref{fig:ratios} shows the maps of the obtained column densities: 
the H$_2$CO column density is $\sim 1.2-3.8 \times 10^{13}$ \cms\, in the disk and $0.2-1.2 \times 10^{13}$ \cms\, in the streamer; 
N(HDCO) is $3-8.5 \times 10^{12}$ \cms\ in the SE and NW disk sides, and $2-3 \times 10^{12}$ \cms\ in the streamer and in the outflow; 
N(D$_2$CO) is $\sim 1-4 \times 10^{12}$ \cms\, in the SE disk side  and $1-2 \times 10^{12}$ \cms\ in the outflow spots along the disk minor axis.


From the obtained column densities maps, we estimate the abundance ratios maps [HDCO]/[H$_2$CO] = $N_{\rm HDCO}/N_{\rm H_2CO}$, [D$_2$CO]/[H$_2$CO] = $N_{\rm D_2CO}/N_{\rm H_2CO}$, and [D$_2$CO]/[HDCO] = $N_{\rm D_2CO}/N_{\rm HDCO}$ (see Fig.~\ref{fig:ratios}).
The deuterium fractionation, [D]/[H], can  be inferred from the abundance ratio between the deuterated isotopologue and the main conformer by taking into account the number of arrangements of the deuterium atoms \citep[e.g., ][]{manigand19}, namely:
\begin{equation}
\frac{{\rm X H}_{n-i} {\rm D}_{i}}{{\rm X H}_n} = \frac{n!}{i! (n-i)!} \left( \frac{\rm D}{\rm H} \right)^i
\end{equation}
Therefore, [D]/[H] = 0.5 $\times$ [HDCO]/[H$_2$CO], and [D]/[H] $= \sqrt{\rm [D_2CO]/[H_2CO]}$.
The deuteration of HDCO in the two disk sides and in the blue- and red-shifted outflow lobes is  [HDCO]/[H$_2$CO] $\sim 0.1-0.3$, which implies [D]/[H] $\sim 5\%-15\%$.
Doubly deuterated formaldehyde is more abundant in the SE disk side, therefore the abundance ratio [D$_2$CO]/[H$_2$CO] is highly asymmetric in the disk:  [D$_2$CO]/[H$_2$CO] $\sim 0.1$ ([D]/[H]$\sim32\%$) in the SE disk, while from the integrated spectra we estimate a factor 2 lower deuteration in the NW disk side ([D$_2$CO]/[H$_2$CO]$\sim 0.05$, implying [D]/[H]$\sim 22\%$).
The [D$_2$CO]/[HDCO] abundance ratio varies between 0.3 and 0.7 in the SE disk side and in the outflow lobes.
As shown in Fig.~\ref{fig:deut}, the [HDCO]/[H$_2$CO] and [D$_2$CO]/[H$_2$CO] abundance ratios estimated in the Class I disk of IRS~63 are, on average, at the upper edge of the range of values estimated in prestellar cores \citep{chacon-tanarro19,bacmann03}, in Class 0 \citep{parise06,watanabe12,persson18,manigand20,evans23}
and Class I \citep{bianchi17a,mercimek22} protostars, and in agreement with the only available upper limit obtained in a disk (Oph IRS48, \citealt{vandermarel21b,brunken22}).
Moreover, the [HDCO]/[H$_2$CO] abundance ratios in the outflow shocks driven by IRS~63 ($\sim 0.1-0.3$) are in agreement with that estimated in the L1157-B1 protostellar shock ($\sim 0.1$, \citealt{fontani14b}).

\section{Discussion}
\label{sect:discussion}

\subsection{{\rm D$_2$CO} in the accretion and outflow shocks}

Although H$_2$CO has been routinely detected in disks \citep[e.g., ][]{garufi21,oberg21}, in outflows \citep[e.g., ][]{tychoniec19,evans23}, and also in a streamer \citep{valdivia-mena22}, this is the first time that singly and doubly deuterated formaldehyde are detected in a planet-forming disk and in the streamer that is feeding it. 
Interestingly, our observations show the following features in the distribution and abundance of HDCO and D$_2$CO in IRS~63:  (i) the spatial distribution of HDCO is similar to that of H$_2$CO in the disk and in the streamer, pointing to a similar chemical origin of the two species; (ii) instead, D$_2$CO emission is detected only in the SE disk side where the streamer hits the disk and in the two spots along the outflow direction, hence in shocked regions; (iii) the [D$_2$CO]/[H$_2$CO] abundance ratio in the SE disk side hit by the streamer is $\sim 0.1$, i.e. at the upper edge of the range of values estimated for prestellar cores and Class 0 and I protostars (see Fig.~\ref{fig:deut}).

The observed spatial distribution and abundance of D$_2$CO could be due to the release of D$_2$CO from the icy mantles of grains in the accretion shock occurring when the streamer impacts onto the disk, as well as in the shocks along the outflow.
The release of molecular species from the ices in the accretion shock at the streamer-disk interface have been first observed by \citet{garufi22}, through enhanced SO and SO$_2$ emission  in the shocked disk region of HL Tau and DG Tau, then by \citet{bianchi23b}, through HDO emission  in SVS13-A.
\citet{flores23} showed that the streamer feeds the disk of IRS~63 at a rate of $\sim 10^{-6}$ \msolyr, two orders of magnitude larger than the disk-to-star mass accretion rate. This suggests that the streamer may have a deep impact not only on the mass and kinematics of the disk, but also on its chemical composition.

In the following section, we examine the different routes for the formation of HDCO and D$_2$CO, either in gas-phase or on the grains, and use the different spatial distribution of these molecules to put constraints on their chemical origin and to verify the above proposed scenario of D$_2$CO-rich ices released in shocked regions.




\subsection{{\rm HDCO} and {\rm D$_2$CO} chemistry}

\begin{figure}
    \centering
    \includegraphics[width=9.cm]{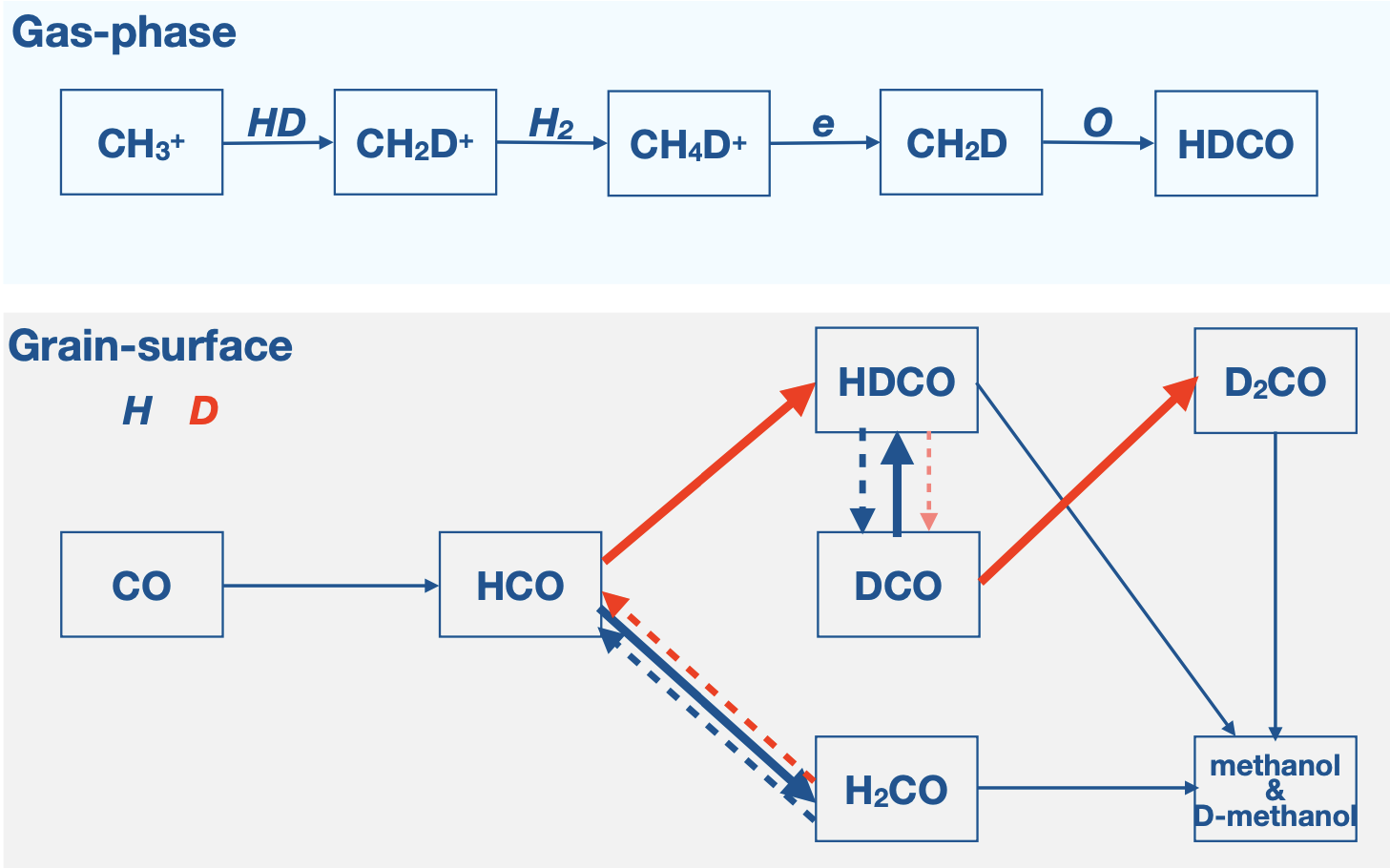}
    \caption{Major routes of formation and destruction of HDCO in the gas-phase (upper panel) and HDCO and D$_2$CO on the grain surfaces (lower panel). 
    In the latter, arrows identify the most important reactions involving H (blue) and D (red) atoms landing on the surface: direct additions (solid lines) and abstraction (dashed lines). 
    Thick lines represent barrierless reactions while thin lines reactions presenting barriers (see text).}
    \label{fig:D2CO-scheme}
\end{figure}

Formaldehyde can be formed both in the gas-phase (mainly through the reaction CH$_3$+O), and on the surfaces of the grain mantles, through hydrogenation of CO ices \citep[e.g., ][]{willacy09}.

The scheme of the two routes forming HDCO is schematically summarized in Fig. \ref{fig:D2CO-scheme}.
In the gas-phase, deuterated formaldehyde is formed mainly via a chain of reactions starting from the CH$_3^+$ ions which, after reacting with HD, forms CH$_2$D$^+$, which eventually leads to HDCO \citep[e.g., ][]{roueff07,roberts07}.
Since the inverse of the first reaction has an exothermicity of about 650 K \citep{Roueff2013-CH2D+}, at temperatures lower than about 70--80 K it does not occur, enhancing the CH$_2$D$^+$/CH$_3+$ ratio, analogously to the H$_2$D$^+$/H$_3+$ ratio at even lower temperatures ($\leq 30-40$ K) \citep{ceccarelli14}.
Theoretical models predict [HDCO]/[H$_2$CO] ratios up to 0.2 \citep{Bergman2011-d2co, Roueff2013-CH2D+}, comparable to the values observed in the disk. 

On the other hand, gas-phase reactions are unable to reproduce the large [D$_2$CO]/[HDCO] ratio inferred in the region where the streamer falls onto the disk (up to $\sim 0.7$).
In this case, the most reasonable explanation is that the large measured abundance of D$_2$CO is due to its release in the gas-phase from the grain mantles, where D$_2$CO was formed in the prestellar phase or recently in the disk itself. 
Fig. \ref{fig:D2CO-scheme} shows the scheme for the formation of D$_2$CO on the grain mantles, which occurs thanks to addition and abstraction reactions of H and D to and from frozen CO.
It is worth noticing the crucial importance of the abstraction reactions in the formation of D$_2$CO.
This was first noticed by experimental works \citep[e.g.,][]{Hidaka2009-D2COeDMethanol} and astrochemical models \citep{Taquet2012-Deuteration, Aikawa2012-Dchemistry}.
Later, highly accurate theoretical calculations have confirmed the importance of abstraction reactions on simulated amorphous water surfaces (AWS) and the Kinetic Isotope Effect (KIE) in the reactions with activation barriers \citep{Song17}. 
However, at our best knowledge, theoretical calculations have been only carried out for the H or D + H$_2$CO (or HDCO) $\rightarrow$ HCO + H$_2$ or (HD), leaving the whole quantitatively reaction scheme incomplete. 

That said, D$_2$CO is formed mainly from the DCO + D reaction and, thus, it uniquely depends on the DCO abundance, which is set by the HDCO abstraction reactions against the addition ones.
Interestingly, the H abstraction of HDCO is always favored against the D abstraction, so that even the H irradiation plays in favor of augmenting the DCO abundance and, consequently, the D$_2$CO one.
Based on the above discussed grain surface chemistry, we suggest that the ices in the disk, as well as those delivered by the streamer, were exposed to an important flux of D atoms that formed HDCO, which was subsequently converted into D$_2$CO due to the flux of H atoms. This 
caused an enhancement of the D$_2$CO abundance (with respect to HDCO and H$_2$CO) on the ices. 
The above described enhancement of  D$_2$CO on the ices, may have occurred in the  cold prestellar phase, to be then inherited by the disk at the time of its formation as well as through the streamer. In facts, previous studies indicate 
an accumulation of dust with thick icy mantles in the central region of prestellar cores and almost complete freeze-out, which favors the formation of deuterated molecules on the ices \citep{ceccarelli14,caselli22}, and this is confirmed by the high [HDCO]/[H$_2$CO] and [D$_2$CO]/[H$_2$CO] abundance ratios inferred in prestellar cores (see Fig.~\ref{fig:deut}). On the other hand, the same grain chemistry may also occur in the cold disk midplane.
In both cases, the D$_2$CO enriched ice composition is released in the gas-phase in the shocks driven by the accretion streamer and the outflow.

To summarize, in light of the formation routes discussed above, we conclude that:
(i) the widespread HDCO and H$_2$CO emission in the disk and in the streamer (Fig. \ref{fig:mom}) can be explained by gas-phase reactions occurring in the warm disk molecular layers and in the streamer at temperatures of $40-50$ K, where the main formation routes of H$_2$CO is CH$_3$ + O, and that of HDCO is CH$_2$D + O (through the scheme presented in Fig. \ref{fig:D2CO-scheme}).
Therefore, both H$_2$CO and HDCO probe material which is (at least partially) reprocessed in the disk and in the streamer;
(ii) D$_2$CO, which is detected only in  shocks (Fig.~\ref{fig:mom}), probes the composition of the ices, which is set by grain surface chemistry occurring at low temperatures, i.e. at the prestellar stage and in the disk midplane.




\section{Conclusions}
ALMA observations of the Class I source IRS~63 in the context of the FAUST ALMA LP allowed the first detection of singly and doubly deuterated formaldehyde in a planet-forming disk and in the streamer that is feeding it. 
A strong chemical asymmetry between the two disk sides is observed. The spatial distribution of H$_2$CO and HDCO in the disk and in the streamer is similar, while D$_2$CO is detected only: (i) in the South-East disk side, where the streamer impacts on the disk, as well as (ii) in two spots along the outflow.
The estimated deuterium fraction, [D]/[H], is $\sim5\%-15\%$ in HDCO and $\sim22-32\%$ in D$_2$CO.
The spatial distribution and abundance of the detected species suggest that HDCO is mainly a gas-phase product (starting from the reaction CH$_3^+$ + HD), while D$_2$CO is formed on the ices, at the prestellar phase and/or in the disk itself, through addition and abstraction reactions of H and D to and from frozen CO \citep[e.g., ][]{Taquet2012-Deuteration}. This explains why D$_2$CO is observed only in shocked regions, either in the accretion shock produced by the streamer or in the outflow shocks, where the ices are sputtered releasing D$_2$CO in gas-phase.
Our results indicate that streamers play a key role in setting the initial conditions for planet formation, as they can alter the disk physical and chemical properties.

\begin{acknowledgements}
This paper makes use of the following ALMA data:
ADS/JAO.ALMA\#2018.1.01205.L (PI: S. Yamamoto). ALMA
is a partnership of the ESO (representing its member states), the NSF (USA) and NINS (Japan), together with the NRC (Canada) and the NSC and ASIAA (Taiwan), in cooperation with the Republic of Chile. The Joint ALMA Observatory is operated by the ESO, the AUI/NRAO, and the NAOJ.
The project has received funding from the EC H2020 research and innovation programme for the project "Astro-Chemical Origins” (ACO, No 811312). LP, ClCo, and GS acknowledge the PRIN-MUR 2020  BEYOND-2p (Astrochemistry beyond the second period elements, Prot. 2020AFB3FX), the project ASI-Astrobiologia 2023 MIGLIORA (Modeling Chemical Complexity, F83C23000800005), the INAF-GO 2023 fundings PROTO-SKA (Exploiting ALMA data to study planet forming disks: preparing the advent of SKA, C13C23000770005), the INAF Mini-Grant 2022
“Chemical Origins” (PI: L. Podio), and the INAF-Minigrant 2023 TRIESTE (“TRacing the chemIcal hEritage of our originS: from proTostars to plan- Ets”; PI: G. Sabatini).
LP, and ClCo also acknowledge financial support under the National Recovery and Resilience Plan (NRRP), Mission 4, Component 2, Investment 1.1, Call for tender No. 104 published on 2.2.2022 by the Italian Ministry of University and Research (MUR), funded by the European Union – NextGenerationEU– Project Title 2022JC2Y93 Chemical Origins: linking the fossil composition of the Solar System with the chemistry of protoplanetary disks – CUP J53D23001600006 - Grant Assignment Decree No. 962 adopted on 30.06.2023 by the Italian Ministry of Ministry of University and Research (MUR).  M.B. acknowledges support from the European Research Council (ERC) Advanced Grant MOPPEX 833460.
S.N. is grateful for support from MEXT/JSPS Grants-in-Aid for Scientific Research (KAKENHI) grant Nos. JP20H05844, JP20H05845. and JP20H05645.
E.B aknowledges the Deutsche Forschungsgemeinschaft (DFG, German Research Foundation) under Germany Excellence Strategy – EXC 2094 – 390783311.
L.L. acknowledges the support of DGAPA PAPIIT grants IN108324 and IN112820 and CONACyT-CF grant 263356.
This project has received funding from the European Research Council (ERC) under the European Union Horizon Europe research and innovation program (grant agreement No. 101042275, project Stellar-MADE).
SBC was supported by the NASA Planetary Science Division Internal Scientist Funding Program through the Fundamental Laboratory Research work package (FLaRe). 
D.J.\ is supported by NRC Canada and by an NSERC Discovery Grant.
I.J-.S acknowledges funding from grant No. PID2022-136814NB-I00 funded by MICIU/AEI/ 10.13039/501100011033 and by “ERDF/EU”.
Y.A acknowledges the support by NAOJ ALMA Scientific Research Grant code 2019-13B.
\end{acknowledgements}


\bibliographystyle{aa} 
\bibliography{mybibtex.bib} 


\begin{appendix}

\section{Channel maps}

\begin{figure*}[ht]
    \centering
     \includegraphics[width=\linewidth]{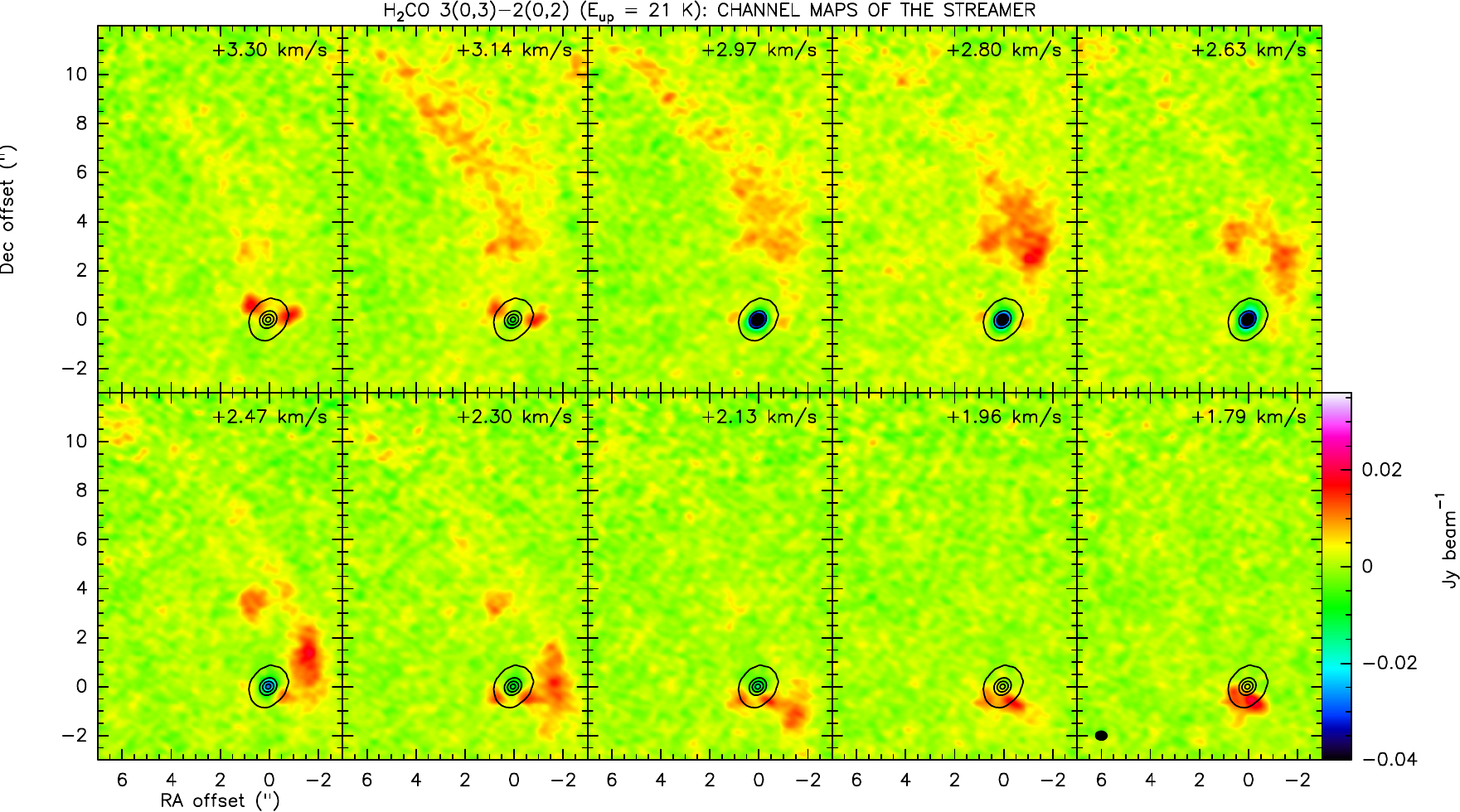}
    \caption{Zoom-out of H$_2$CO 218.22 GHz in the channels at $V_{\rm LSR} = [+1.79, +3.3]$ \kms\, to show the emission by the large scale streamer. The black contours show the continuum emission at 1.3~mm (contours from $10\sigma$ with steps of $500\sigma$, with $\sigma = 0.7$ mJy\,beam$^{-1}$). In each panel the velocity, $V_{\rm LSR}$, is indicated on the top-right (channel width of 0.168 \kms\, channel). The beam and the wedge of the color scale are drawn in the last channel.}
    \label{fig:h2co_streamer_channels}
\end{figure*}

\begin{figure*}[ht]
    \centering
    \includegraphics[width=\linewidth]{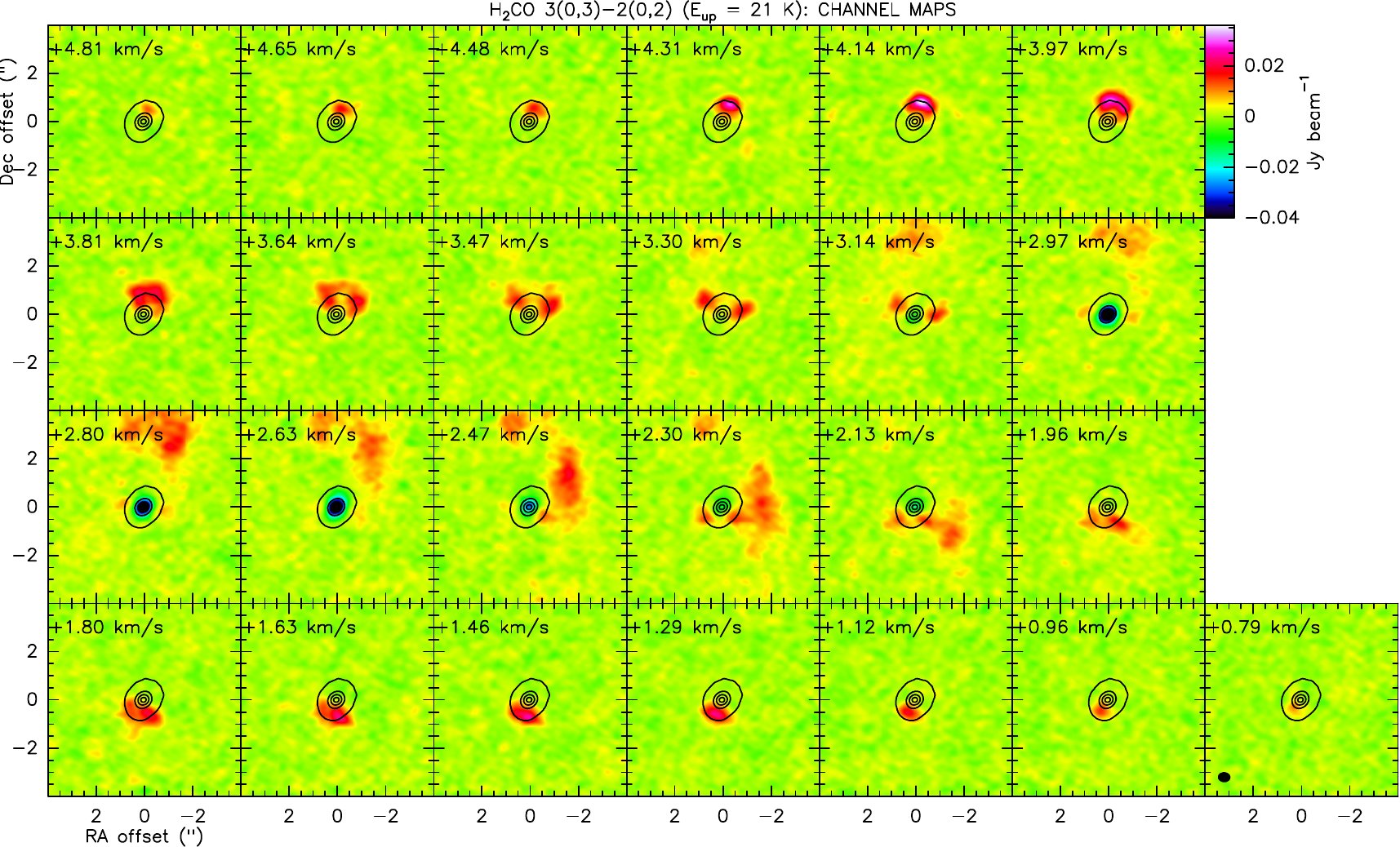}
    \caption{Channel maps of H$_2$CO 218.22 GHz emission towards IRS~63. The black contours show the continuum emission at 1.3~mm (contours from $10\sigma$ with steps of $500\sigma$, with $\sigma = 0.7$ mJy\,beam$^{-1}$). In each panel the velocity, $V_{\rm LSR}$, is indicated on the top-left (channel width of 0.168 \kms\, channel). The wedge and the beam are drawn in the last channel of the first and last row, respectively.}
    \label{fig:h2co_channels}
\end{figure*}

\begin{figure*}[ht]
    \centering
    \includegraphics[width=12.cm]{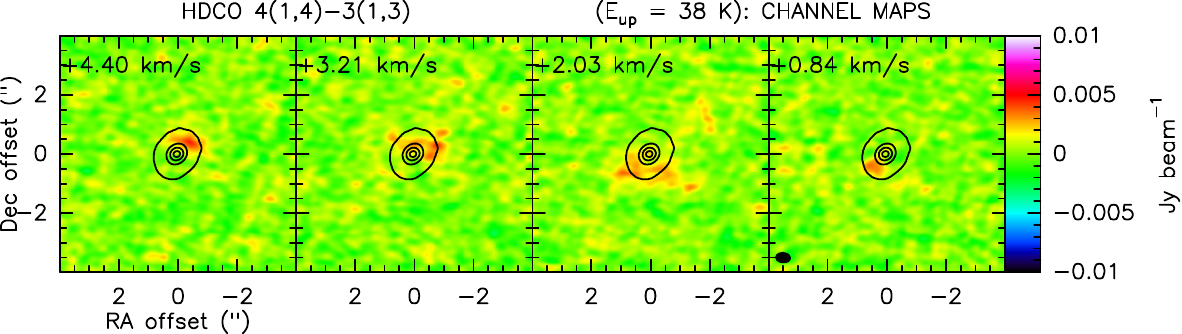}
    \caption{Channel maps of HDCO 246.92 GHz emission towards IRS~63. The black contours show the continuum emission at 1.3~mm (contours from $10\sigma$ with steps of $500\sigma$, with $\sigma = 0.7$ mJy\,beam$^{-1}$). The scale of the wedge is in Jy beam$^{-1}$. In each panel the velocity, $V_{\rm LSR}$, is indicated on the top-left (channel width of 1.19 \kms\, channel). The beam and the wedge of the color scale are drawn in the last channel.}
    \label{fig:hdco_channels}
\end{figure*}

\begin{figure*}[ht]
    \centering
    \includegraphics[width=\linewidth]{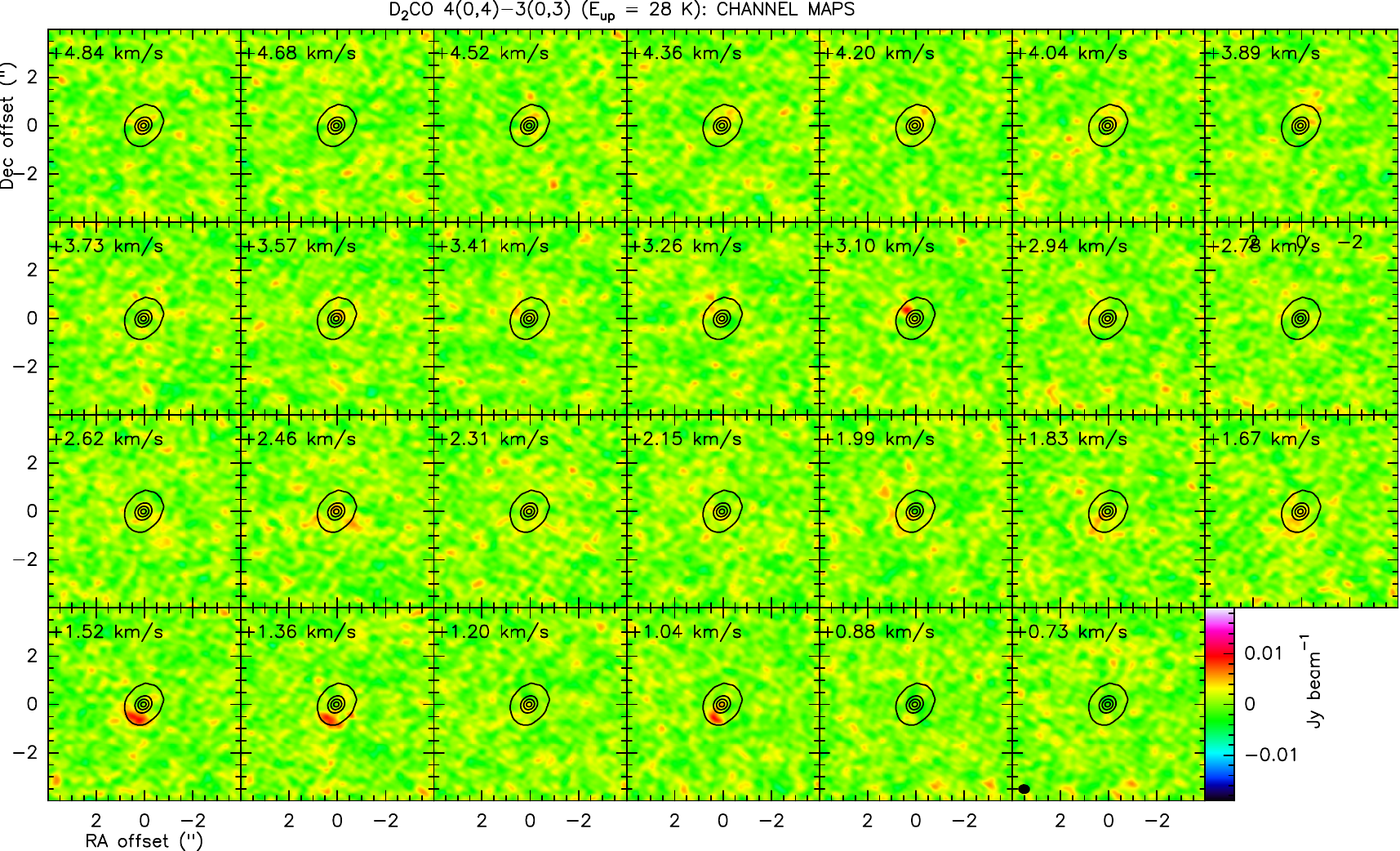}
    \caption{Channel maps of D$_2$CO 231.41 GHz emission towards IRS~63. The black contours show the continuum emission at 1.3~mm (contours from $10\sigma$ with steps of $500\sigma$, with $\sigma = 0.7$ mJy\,beam$^{-1}$). The scale of the wedge is in Jy beam$^{-1}$. In each panel the velocity, $V_{\rm LSR}$, is indicated on the top-left (channel width of 0.158 \kms\, channel). The beam and the wedge of the color scale are drawn in the last channel.}
    \label{fig:d2co_channels}
\end{figure*}

Channel maps of H$_2$CO 218.22 GHz, HDCO 246.92 GHz, and D$_2$CO 231.41 GHz towards IRS~63 are shown in Figs. \ref{fig:h2co_channels}, \ref{fig:hdco_channels}, and \ref{fig:d2co_channels}.
Channels maps of H$_2$CO on a larger area of $10"\times15"$ are shown in Fig. \ref{fig:h2co_streamer_channels}
to show emission from the large scale streamer. 
The channel width is different for each line, as reported in Table~\ref{tab:lines}.

\section{Integrated spectra}

\begin{figure*}[ht]
    \centering
    \includegraphics[width=\linewidth]{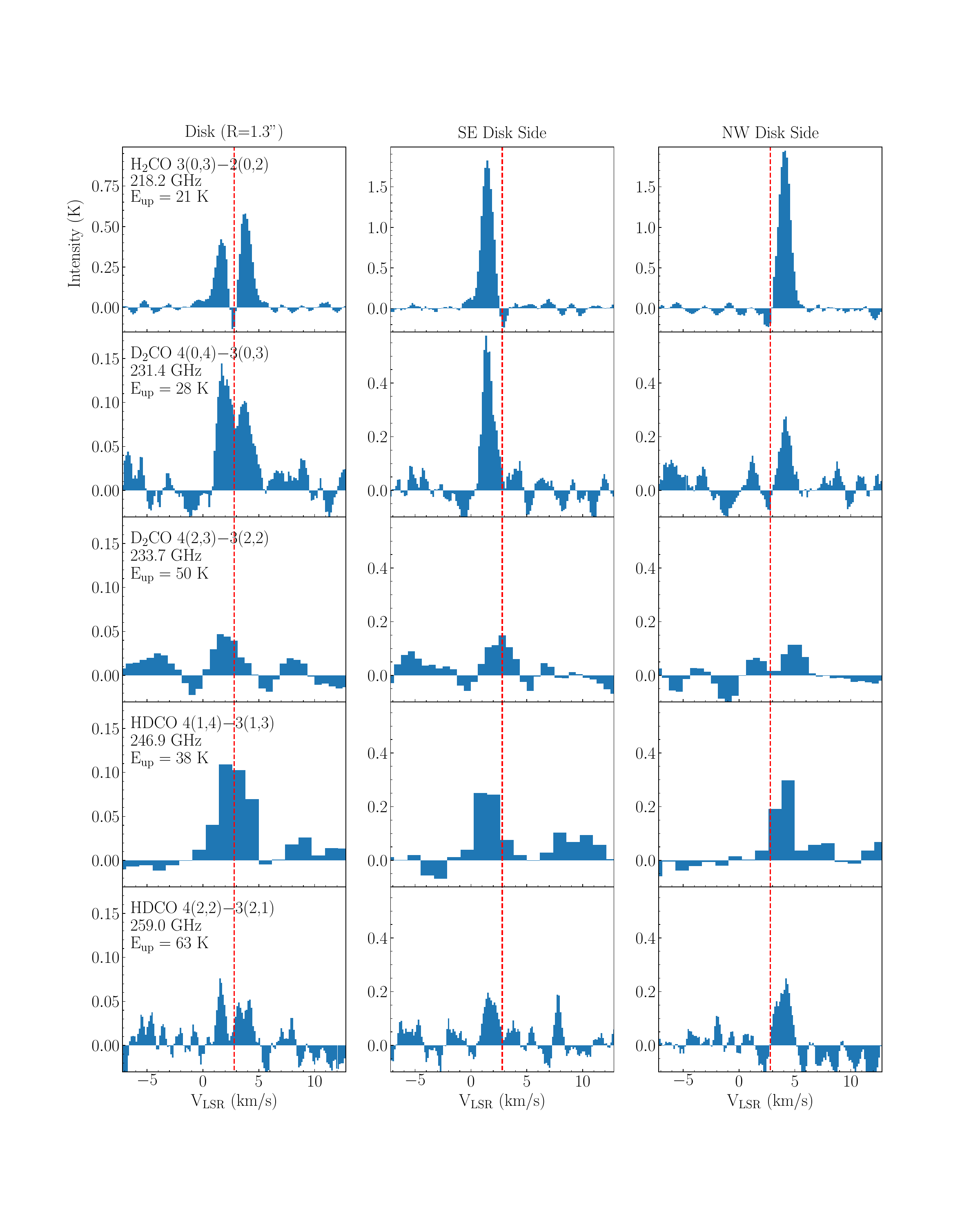}
    \caption{Spectra of the H$_2$CO, HDCO, and D$_2$CO lines integrated: on the entire disk (out to a radius of $1\farcs3$, {\it 1st column}), on the SE disk side ({\it 2nd column}), on the NW disk side ({\it 3rd column}). The red dashed line indicates the systemic velocity, $V_{\rm LSR} = +2.8$ \kms. The line transition, frequency, and upper level energy are labeled on the upper-left of the panels in the first column.}
    \label{fig:spec}
\end{figure*}

In Fig.~\ref{fig:spec} we show the spectra of H$_2$CO, HDCO, and D$_2$CO transitions integrated on the entire disk (out to a radius of $\sim 1\farcs3$, i.e. 187 au), on  the SE and NW disk side.

\section{Abundance and [D]/[H] ratios from prestellar cores to disks}

\begin{figure}[ht]
   \centering
   \includegraphics[width=9.cm]{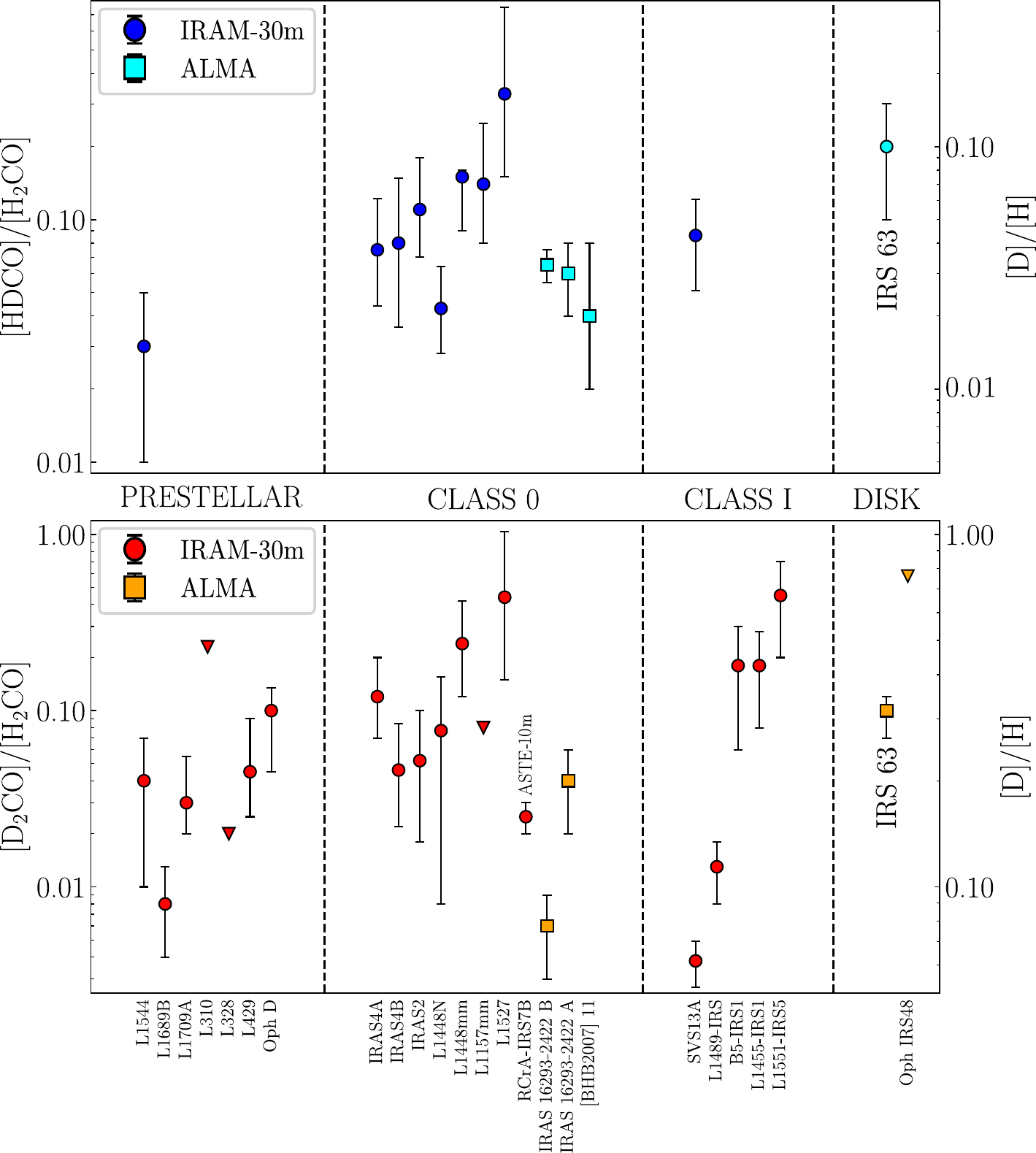}
    \caption{Abundance ratios [HDCO]/[H$_2$CO] ({\it upper panel}), and [D$_2$CO]/[H$_2$CO] ({\it lower panel}) estimated in prestellar cores \citep{chacon-tanarro19,bacmann03}, in Class 0 \citep{parise06,watanabe12,persson18,manigand20,evans23} and Class I \citep{bianchi17a,mercimek22} protostars, and in disks \citep{vandermarel21b,brunken22}, are compared with the values estimated in the disk of IRS~63 (this work).
    Blue and red circles indicate estimates obtained with IRAM-30m observations (with the exception of RCrA-IRS7B which is observed with ASTE-10m), while cyan and orange squares those obtained with ALMA. Upper limits are marked with triangles. 
    In both panels the right y-axes indicate the corresponding [D]/[H] ratios. 
    The name of the sources are labeled in the bottom panel. 
    }
    \label{fig:deut}
\end{figure}

Figure \ref{fig:deut} compares the [HDCO]/[H$_2$CO] and [D$_2$CO]/[H$_2$CO] abundance ratios, and corresponding [D]/[H], in the Class I disk of IRS~63 with the values previuosly estimated for sources at different evolutionary stages, from prestellar cores, to Class 0 and I protostars and disks. The corresponding source names are labeled in the figure and the references are given in the caption. 

\end{appendix}

\end{document}